\documentclass[
superscriptaddress,
nofootinbib,
nobibnotes,
notitlepage,
amsmath,amssymb,
aps,
twocolumn,
prd,
10pt
]{revtex4-2}

\newcommand\farcm{\hbox{$.\mkern-4mu^\prime$}}

\newcommand\be{\begin{equation}}
\newcommand\ee{\end{equation}}

\newcommand\comments[1]{}
\usepackage{amsmath}
\usepackage{amssymb}
\usepackage{verbatim}
\usepackage{tabularx}
\usepackage{makecell}

\usepackage[T1]{fontenc}
\usepackage{aas_macros}
\usepackage{graphicx}
\usepackage{dcolumn}
\usepackage{lipsum, babel}
\usepackage{orcidlink}
\usepackage{booktabs}

\newcommand{\rsigmafullsimple}{\ensuremath{0.068}}
\newcommand{\Deightyonefiftyghzconstraintfullsimple}{\ensuremath{0.037^{+0.009}_{-0.008}}}
\newcommand{\betagalconstraintfullsimple}{\ensuremath{1.670^{+0.269}_{-0.254}}}
\newcommand{\alphagalconstraintfullsimple}{\ensuremath{-0.956^{+0.196}_{-0.158}}}
\newcommand{\Dfivehundredninetyfiveghzsyncpoissonconstraintfullsimple}{\ensuremath{0.036^{+0.010}_{-0.009}}}
\newcommand{\rlimitfullsimple}{\ensuremath{0.25}}

\newcommand{\rsigmabksimple}{\ensuremath{0.067}}
\newcommand{\Deightyonefiftyghzconstraintbksimple}{\ensuremath{0.013^{+0.006}_{-0.004}}}
\newcommand{\betagalconstraintbksimple}{\ensuremath{2.197^{+0.484}_{-0.545}}}
\newcommand{\alphagalconstraintbksimple}{\ensuremath{-0.729^{+0.293}_{-0.290}}}
\newcommand{\Dfivehundredninetyfiveghzsyncpoissonconstraintbksimple}{\ensuremath{0.011^{+0.010}_{-0.007}}}
\newcommand{\rlimitbksimple}{\ensuremath{0.25}}

\graphicspath{{./}{figures/}}
 
\begin{document}


\title{Constraints on Inflationary Gravitational Waves with Two Years of SPT-3G Data}
\author{J.~A.~Zebrowski \,\orcidlink{0000-0003-2375-0229}}
\email{j.z@uchicago.edu}
\altaffiliation{NASA Einstein Fellow}
\affiliation{Kavli Institute for Cosmological Physics, University of Chicago, 5640 South Ellis Avenue, Chicago, IL, 60637, USA}
\affiliation{Department of Astronomy and Astrophysics, University of Chicago, 5640 South Ellis Avenue, Chicago, IL, 60637, USA}
\affiliation{Fermi National Accelerator Laboratory, MS209, P.O. Box 500, Batavia, IL, 60510, USA}

\author{C.~L.~Reichardt \,\orcidlink{0000-0003-2226-9169}}
\affiliation{School of Physics, University of Melbourne, Parkville, VIC 3010, Australia}

\author{A.~J.~Anderson\,\orcidlink{0000-0002-4435-4623}}
\affiliation{Fermi National Accelerator Laboratory, MS209, P.O. Box 500, Batavia, IL, 60510, USA}
\affiliation{Kavli Institute for Cosmological Physics, University of Chicago, 5640 South Ellis Avenue, Chicago, IL, 60637, USA}
\affiliation{Department of Astronomy and Astrophysics, University of Chicago, 5640 South Ellis Avenue, Chicago, IL, 60637, USA}
\author{B.~Ansarinejad}
\affiliation{School of Physics, University of Melbourne, Parkville, VIC 3010, Australia}
\author{M.~Archipley\,\orcidlink{0000-0002-0517-9842}}
\affiliation{Department of Astronomy and Astrophysics, University of Chicago, 5640 South Ellis Avenue, Chicago, IL, 60637, USA}
\affiliation{Kavli Institute for Cosmological Physics, University of Chicago, 5640 South Ellis Avenue, Chicago, IL, 60637, USA}
\author{L.~Balkenhol\,\orcidlink{0000-0001-6899-1873}}
\affiliation{Sorbonne Universit\'e, CNRS, UMR 7095, Institut d'Astrophysique de Paris, 98 bis bd Arago, 75014 Paris, France}
\author{P.~S.~Barry}
\affiliation{School of Physics and Astronomy, Cardiff University, Cardiff, CF24 3AA, UK}
\author{K.~Benabed}
\affiliation{Sorbonne Universit\'e, CNRS, UMR 7095, Institut d'Astrophysique de Paris, 98 bis bd Arago, 75014 Paris, France}
\author{A.~N.~Bender\,\orcidlink{0000-0001-5868-0748}}
\affiliation{High-Energy Physics Division, Argonne National Laboratory, 9700 South Cass Avenue., Lemont, IL, 60439, USA}
\affiliation{Kavli Institute for Cosmological Physics, University of Chicago, 5640 South Ellis Avenue, Chicago, IL, 60637, USA}
\affiliation{Department of Astronomy and Astrophysics, University of Chicago, 5640 South Ellis Avenue, Chicago, IL, 60637, USA}
\author{B.~A.~Benson\,\orcidlink{0000-0002-5108-6823}}
\affiliation{Fermi National Accelerator Laboratory, MS209, P.O. Box 500, Batavia, IL, 60510, USA}
\affiliation{Kavli Institute for Cosmological Physics, University of Chicago, 5640 South Ellis Avenue, Chicago, IL, 60637, USA}
\affiliation{Department of Astronomy and Astrophysics, University of Chicago, 5640 South Ellis Avenue, Chicago, IL, 60637, USA}
\author{F.~Bianchini\,\orcidlink{0000-0003-4847-3483}}
\affiliation{Kavli Institute for Particle Astrophysics and Cosmology, Stanford University, 452 Lomita Mall, Stanford, CA, 94305, USA}
\affiliation{Department of Physics, Stanford University, 382 Via Pueblo Mall, Stanford, CA, 94305, USA}
\affiliation{SLAC National Accelerator Laboratory, 2575 Sand Hill Road, Menlo Park, CA, 94025, USA}
\author{L.~E.~Bleem\,\orcidlink{0000-0001-7665-5079}}
\affiliation{High-Energy Physics Division, Argonne National Laboratory, 9700 South Cass Avenue., Lemont, IL, 60439, USA}
\affiliation{Kavli Institute for Cosmological Physics, University of Chicago, 5640 South Ellis Avenue, Chicago, IL, 60637, USA}
\affiliation{Department of Astronomy and Astrophysics, University of Chicago, 5640 South Ellis Avenue, Chicago, IL, 60637, USA}
\author{F.~R.~Bouchet\,\orcidlink{0000-0002-8051-2924}}
\affiliation{Sorbonne Universit\'e, CNRS, UMR 7095, Institut d'Astrophysique de Paris, 98 bis bd Arago, 75014 Paris, France}
\author{L.~Bryant}
\affiliation{Enrico Fermi Institute, University of Chicago, 5640 South Ellis Avenue, Chicago, IL, 60637, USA}
\author{E.~Camphuis\,\orcidlink{0000-0003-3483-8461}}
\affiliation{Sorbonne Universit\'e, CNRS, UMR 7095, Institut d'Astrophysique de Paris, 98 bis bd Arago, 75014 Paris, France}
\author{J.~E.~Carlstrom\,\orcidlink{0000-0002-2044-7665}}
\affiliation{Kavli Institute for Cosmological Physics, University of Chicago, 5640 South Ellis Avenue, Chicago, IL, 60637, USA}
\affiliation{Enrico Fermi Institute, University of Chicago, 5640 South Ellis Avenue, Chicago, IL, 60637, USA}
\affiliation{Department of Physics, University of Chicago, 5640 South Ellis Avenue, Chicago, IL, 60637, USA}
\affiliation{High-Energy Physics Division, Argonne National Laboratory, 9700 South Cass Avenue., Lemont, IL, 60439, USA}
\affiliation{Department of Astronomy and Astrophysics, University of Chicago, 5640 South Ellis Avenue, Chicago, IL, 60637, USA}
\author{C.~L.~Chang}
\affiliation{High-Energy Physics Division, Argonne National Laboratory, 9700 South Cass Avenue., Lemont, IL, 60439, USA}
\affiliation{Kavli Institute for Cosmological Physics, University of Chicago, 5640 South Ellis Avenue, Chicago, IL, 60637, USA}
\affiliation{Department of Astronomy and Astrophysics, University of Chicago, 5640 South Ellis Avenue, Chicago, IL, 60637, USA}
\author{P.~Chaubal}
\affiliation{School of Physics, University of Melbourne, Parkville, VIC 3010, Australia}
\author{P.~M.~Chichura\,\orcidlink{0000-0002-5397-9035}}
\affiliation{Department of Physics, University of Chicago, 5640 South Ellis Avenue, Chicago, IL, 60637, USA}
\affiliation{Kavli Institute for Cosmological Physics, University of Chicago, 5640 South Ellis Avenue, Chicago, IL, 60637, USA}
\author{A.~Chokshi}
\affiliation{University of Chicago, 5640 South Ellis Avenue, Chicago, IL, 60637, USA}
\author{T.-L.~Chou\,\orcidlink{0000-0002-3091-8790}}
\affiliation{Department of Astronomy and Astrophysics, University of Chicago, 5640 South Ellis Avenue, Chicago, IL, 60637, USA}
\affiliation{Kavli Institute for Cosmological Physics, University of Chicago, 5640 South Ellis Avenue, Chicago, IL, 60637, USA}
\author{A.~Coerver}
\affiliation{Department of Physics, University of California, Berkeley, CA, 94720, USA}
\author{T.~M.~Crawford\,\orcidlink{0000-0001-9000-5013}}
\affiliation{Department of Astronomy and Astrophysics, University of Chicago, 5640 South Ellis Avenue, Chicago, IL, 60637, USA}
\affiliation{Kavli Institute for Cosmological Physics, University of Chicago, 5640 South Ellis Avenue, Chicago, IL, 60637, USA}
\author{C.~Daley\,\orcidlink{0000-0002-3760-2086}}
\affiliation{Universit\'e Paris-Saclay, Universit\'e Paris Cit\'e, CEA, CNRS, AIM, 91191, Gif-sur-Yvette, France}
\affiliation{Department of Astronomy, University of Illinois Urbana-Champaign, 1002 West Green Street, Urbana, IL, 61801, USA}
\author{T.~de~Haan}
\affiliation{High Energy Accelerator Research Organization (KEK), Tsukuba, Ibaraki 305-0801, Japan}
\author{K.~R.~Dibert}
\affiliation{Department of Astronomy and Astrophysics, University of Chicago, 5640 South Ellis Avenue, Chicago, IL, 60637, USA}
\affiliation{Kavli Institute for Cosmological Physics, University of Chicago, 5640 South Ellis Avenue, Chicago, IL, 60637, USA}
\author{M.~A.~Dobbs}
\affiliation{Department of Physics and McGill Space Institute, McGill University, 3600 Rue University, Montreal, Quebec H3A 2T8, Canada}
\affiliation{Canadian Institute for Advanced Research, CIFAR Program in Gravity and the Extreme Universe, Toronto, ON, M5G 1Z8, Canada}
\author{M.~Doohan}
\affiliation{School of Physics, University of Melbourne, Parkville, VIC 3010, Australia}
\author{A.~Doussot}
\affiliation{Sorbonne Universit\'e, CNRS, UMR 7095, Institut d'Astrophysique de Paris, 98 bis bd Arago, 75014 Paris, France}
\author{D.~Dutcher\,\orcidlink{0000-0002-9962-2058}}
\affiliation{Joseph Henry Laboratories of Physics, Jadwin Hall, Princeton University, Princeton, NJ 08544, USA}
\author{W.~Everett}
\affiliation{Department of Astrophysical and Planetary Sciences, University of Colorado, Boulder, CO, 80309, USA}
\author{C.~Feng}
\affiliation{Department of Physics, University of Illinois Urbana-Champaign, 1110 West Green Street, Urbana, IL, 61801, USA}
\author{K.~R.~Ferguson\,\orcidlink{0000-0002-4928-8813}}
\affiliation{Department of Physics and Astronomy, University of California, Los Angeles, CA, 90095, USA}
\affiliation{Department of Physics and Astronomy, Michigan State University, East Lansing, MI 48824, USA}
\author{K.~Fichman}
\affiliation{Department of Physics, University of Chicago, 5640 South Ellis Avenue, Chicago, IL, 60637, USA}
\affiliation{Kavli Institute for Cosmological Physics, University of Chicago, 5640 South Ellis Avenue, Chicago, IL, 60637, USA}
\author{A.~Foster\,\orcidlink{0000-0002-7145-1824}}
\affiliation{Joseph Henry Laboratories of Physics, Jadwin Hall, Princeton University, Princeton, NJ 08544, USA}
\author{S.~Galli}
\affiliation{Sorbonne Universit\'e, CNRS, UMR 7095, Institut d'Astrophysique de Paris, 98 bis bd Arago, 75014 Paris, France}
\author{A.~E.~Gambrel}
\affiliation{Kavli Institute for Cosmological Physics, University of Chicago, 5640 South Ellis Avenue, Chicago, IL, 60637, USA}
\author{R.~W.~Gardner}
\affiliation{Enrico Fermi Institute, University of Chicago, 5640 South Ellis Avenue, Chicago, IL, 60637, USA}
\author{F.~Ge}
\affiliation{Kavli Institute for Particle Astrophysics and Cosmology, Stanford University, 452 Lomita Mall, Stanford, CA, 94305, USA}
\affiliation{Department of Physics, Stanford University, 382 Via Pueblo Mall, Stanford, CA, 94305, USA}
\affiliation{Department of Physics \& Astronomy, University of California, One Shields Avenue, Davis, CA 95616, USA}
\author{N.~Goeckner-Wald}
\affiliation{Department of Physics, Stanford University, 382 Via Pueblo Mall, Stanford, CA, 94305, USA}
\affiliation{Kavli Institute for Particle Astrophysics and Cosmology, Stanford University, 452 Lomita Mall, Stanford, CA, 94305, USA}
\author{R.~Gualtieri\,\orcidlink{0000-0003-4245-2315}}
\affiliation{High-Energy Physics Division, Argonne National Laboratory, 9700 South Cass Avenue., Lemont, IL, 60439, USA}
\affiliation{Department of Physics and Astronomy, Northwestern University, 633 Clark St, Evanston, IL, 60208, USA}
\author{F.~Guidi}
\affiliation{Sorbonne Universit\'e, CNRS, UMR 7095, Institut d'Astrophysique de Paris, 98 bis bd Arago, 75014 Paris, France}
\author{S.~Guns}
\affiliation{Department of Physics, University of California, Berkeley, CA, 94720, USA}
\author{N.~W.~Halverson}
\affiliation{CASA, Department of Astrophysical and Planetary Sciences, University of Colorado, Boulder, CO, 80309, USA }
\affiliation{Department of Physics, University of Colorado, Boulder, CO, 80309, USA}
\author{E.~Hivon}
\affiliation{Sorbonne Universit\'e, CNRS, UMR 7095, Institut d'Astrophysique de Paris, 98 bis bd Arago, 75014 Paris, France}
\author{G.~P.~Holder\,\orcidlink{0000-0002-0463-6394}}
\affiliation{Department of Physics, University of Illinois Urbana-Champaign, 1110 West Green Street, Urbana, IL, 61801, USA}
\author{W.~L.~Holzapfel}
\affiliation{Department of Physics, University of California, Berkeley, CA, 94720, USA}
\author{J.~C.~Hood}
\affiliation{Kavli Institute for Cosmological Physics, University of Chicago, 5640 South Ellis Avenue, Chicago, IL, 60637, USA}
\author{A.~Hryciuk}
\affiliation{Department of Physics, University of Chicago, 5640 South Ellis Avenue, Chicago, IL, 60637, USA}
\affiliation{Kavli Institute for Cosmological Physics, University of Chicago, 5640 South Ellis Avenue, Chicago, IL, 60637, USA}
\author{N.~Huang}
\affiliation{Department of Physics, University of California, Berkeley, CA, 94720, USA}
\author{F.~K\'eruzor\'e}
\affiliation{High-Energy Physics Division, Argonne National Laboratory, 9700 South Cass Avenue., Lemont, IL, 60439, USA}
\author{A.~R.~Khalife}
\affiliation{Sorbonne Universit\'e, CNRS, UMR 7095, Institut d'Astrophysique de Paris, 98 bis bd Arago, 75014 Paris, France}
\author{L.~Knox}
\affiliation{Department of Physics \& Astronomy, University of California, One Shields Avenue, Davis, CA 95616, USA}
\author{M.~Korman}
\affiliation{Department of Physics, Case Western Reserve University, Cleveland, OH, 44106, USA}
\author{K.~Kornoelje}
\affiliation{Department of Astronomy and Astrophysics, University of Chicago, 5640 South Ellis Avenue, Chicago, IL, 60637, USA}
\affiliation{Kavli Institute for Cosmological Physics, University of Chicago, 5640 South Ellis Avenue, Chicago, IL, 60637, USA}
\author{C.-L.~Kuo}
\affiliation{Kavli Institute for Particle Astrophysics and Cosmology, Stanford University, 452 Lomita Mall, Stanford, CA, 94305, USA}
\affiliation{Department of Physics, Stanford University, 382 Via Pueblo Mall, Stanford, CA, 94305, USA}
\affiliation{SLAC National Accelerator Laboratory, 2575 Sand Hill Road, Menlo Park, CA, 94025, USA}
\author{Y.~Li \,\orcidlink{0000-0002-4820-1122}}
\affiliation{Kavli Institute for Cosmological Physics, University of Chicago, 5640 South Ellis Avenue, Chicago, IL, 60637, USA}
\author{K.~Levy}
\affiliation{School of Physics, University of Melbourne, Parkville, VIC 3010, Australia}
\author{A.~E.~Lowitz\,\orcidlink{0000-0002-4747-4276}}
\affiliation{Kavli Institute for Cosmological Physics, University of Chicago, 5640 South Ellis Avenue, Chicago, IL, 60637, USA}
\author{C.~Lu}
\affiliation{Department of Physics, University of Illinois Urbana-Champaign, 1110 West Green Street, Urbana, IL, 61801, USA}
\author{G.~P.~Lynch\,\orcidlink{0009-0004-3143-1708}}
\affiliation{Department of Physics \& Astronomy, University of California, One Shields Avenue, Davis, CA 95616, USA}
\author{A.~Maniyar}
\affiliation{Kavli Institute for Particle Astrophysics and Cosmology, Stanford University, 452 Lomita Mall, Stanford, CA, 94305, USA}
\affiliation{Department of Physics, Stanford University, 382 Via Pueblo Mall, Stanford, CA, 94305, USA}
\affiliation{SLAC National Accelerator Laboratory, 2575 Sand Hill Road, Menlo Park, CA, 94025, USA}
\author{E.~S.~Martsen}
\affiliation{Department of Astronomy and Astrophysics, University of Chicago, 5640 South Ellis Avenue, Chicago, IL, 60637, USA}
\affiliation{Kavli Institute for Cosmological Physics, University of Chicago, 5640 South Ellis Avenue, Chicago, IL, 60637, USA}
\author{F.~Menanteau}
\affiliation{Department of Astronomy, University of Illinois Urbana-Champaign, 1002 West Green Street, Urbana, IL, 61801, USA}
\affiliation{Center for AstroPhysical Surveys, National Center for Supercomputing Applications, Urbana, IL, 61801, USA}
\author{M.~Millea\,\orcidlink{0000-0001-7317-0551}}
\affiliation{Department of Physics, University of California, Berkeley, CA, 94720, USA}
\author{J.~Montgomery}
\affiliation{Department of Physics and McGill Space Institute, McGill University, 3600 Rue University, Montreal, Quebec H3A 2T8, Canada}
\author{Y.~Nakato}
\affiliation{Department of Physics, Stanford University, 382 Via Pueblo Mall, Stanford, CA, 94305, USA}
\author{T.~Natoli}
\affiliation{Kavli Institute for Cosmological Physics, University of Chicago, 5640 South Ellis Avenue, Chicago, IL, 60637, USA}
\author{G.~I.~Noble\,\orcidlink{0000-0002-5254-243X}}
\affiliation{Dunlap Institute for Astronomy \& Astrophysics, University of Toronto, 50 St. George Street, Toronto, ON, M5S 3H4, Canada}
\affiliation{David A. Dunlap Department of Astronomy \& Astrophysics, University of Toronto, 50 St. George Street, Toronto, ON, M5S 3H4, Canada}
\author{Y.~Omori}
\affiliation{Department of Astronomy and Astrophysics, University of Chicago, 5640 South Ellis Avenue, Chicago, IL, 60637, USA}
\affiliation{Kavli Institute for Cosmological Physics, University of Chicago, 5640 South Ellis Avenue, Chicago, IL, 60637, USA}
\author{A.~Ouellette}
\affiliation{Department of Physics, University of Illinois Urbana-Champaign, 1110 West Green Street, Urbana, IL, 61801, USA}
\author{Z.~Pan\,\orcidlink{0000-0002-6164-9861}}
\affiliation{High-Energy Physics Division, Argonne National Laboratory, 9700 South Cass Avenue., Lemont, IL, 60439, USA}
\affiliation{Kavli Institute for Cosmological Physics, University of Chicago, 5640 South Ellis Avenue, Chicago, IL, 60637, USA}
\affiliation{Department of Physics, University of Chicago, 5640 South Ellis Avenue, Chicago, IL, 60637, USA}
\author{P.~Paschos}
\affiliation{Enrico Fermi Institute, University of Chicago, 5640 South Ellis Avenue, Chicago, IL, 60637, USA}
\author{K.~A.~Phadke\,\orcidlink{0000-0001-7946-557X}}
\affiliation{Department of Astronomy, University of Illinois Urbana-Champaign, 1002 West Green Street, Urbana, IL, 61801, USA}
\affiliation{Center for AstroPhysical Surveys, National Center for Supercomputing Applications, Urbana, IL, 61801, USA}
\author{A.~W.~Pollak}
\affiliation{University of Chicago, 5640 South Ellis Avenue, Chicago, IL, 60637, USA}
\author{K.~Prabhu}
\affiliation{Department of Physics \& Astronomy, University of California, One Shields Avenue, Davis, CA 95616, USA}
\author{W.~Quan}
\affiliation{High-Energy Physics Division, Argonne National Laboratory, 9700 South Cass Avenue., Lemont, IL, 60439, USA}
\affiliation{Department of Physics, University of Chicago, 5640 South Ellis Avenue, Chicago, IL, 60637, USA}
\affiliation{Kavli Institute for Cosmological Physics, University of Chicago, 5640 South Ellis Avenue, Chicago, IL, 60637, USA}
\author{S.~Raghunathan\,\orcidlink{0000-0003-1405-378X}}
\affiliation{Center for AstroPhysical Surveys, National Center for Supercomputing Applications, Urbana, IL, 61801, USA}
\author{M.~Rahimi}
\affiliation{School of Physics, University of Melbourne, Parkville, VIC 3010, Australia}
\author{A.~Rahlin\,\orcidlink{0000-0003-3953-1776}}
\affiliation{Department of Astronomy and Astrophysics, University of Chicago, 5640 South Ellis Avenue, Chicago, IL, 60637, USA}
\affiliation{Kavli Institute for Cosmological Physics, University of Chicago, 5640 South Ellis Avenue, Chicago, IL, 60637, USA}
\author{M.~Rouble}
\affiliation{Department of Physics and McGill Space Institute, McGill University, 3600 Rue University, Montreal, Quebec H3A 2T8, Canada}
\author{J.~E.~Ruhl}
\affiliation{Department of Physics, Case Western Reserve University, Cleveland, OH, 44106, USA}
\author{E.~Schiappucci}
\affiliation{School of Physics, University of Melbourne, Parkville, VIC 3010, Australia}
\author{A.~Simpson}
\affiliation{Department of Astronomy and Astrophysics, University of Chicago, 5640 South Ellis Avenue, Chicago, IL, 60637, USA}
\affiliation{Kavli Institute for Cosmological Physics, University of Chicago, 5640 South Ellis Avenue, Chicago, IL, 60637, USA}
\author{J.~A.~Sobrin\,\orcidlink{0000-0001-6155-5315}}
\affiliation{Fermi National Accelerator Laboratory, MS209, P.O. Box 500, Batavia, IL, 60510, USA}
\affiliation{Kavli Institute for Cosmological Physics, University of Chicago, 5640 South Ellis Avenue, Chicago, IL, 60637, USA}
\author{A.~A.~Stark}
\affiliation{Center for Astrophysics \textbar{} Harvard \& Smithsonian, 60 Garden Street, Cambridge, MA, 02138, USA}
\author{J.~Stephen}
\affiliation{Enrico Fermi Institute, University of Chicago, 5640 South Ellis Avenue, Chicago, IL, 60637, USA}
\author{C.~Tandoi}
\affiliation{Department of Astronomy, University of Illinois Urbana-Champaign, 1002 West Green Street, Urbana, IL, 61801, USA}
\author{B.~Thorne}
\affiliation{Department of Physics \& Astronomy, University of California, One Shields Avenue, Davis, CA 95616, USA}
\author{C.~Trendafilova}
\affiliation{Center for AstroPhysical Surveys, National Center for Supercomputing Applications, Urbana, IL, 61801, USA}
\author{C.~Umilta\,\orcidlink{0000-0002-6805-6188}}
\affiliation{Department of Physics, University of Illinois Urbana-Champaign, 1110 West Green Street, Urbana, IL, 61801, USA}
\author{J.~D.~Vieira\,\orcidlink{0000-0001-7192-3871}}
\affiliation{Department of Astronomy, University of Illinois Urbana-Champaign, 1002 West Green Street, Urbana, IL, 61801, USA}
\affiliation{Department of Physics, University of Illinois Urbana-Champaign, 1110 West Green Street, Urbana, IL, 61801, USA}
\affiliation{Center for AstroPhysical Surveys, National Center for Supercomputing Applications, Urbana, IL, 61801, USA}
\author{A.~G.~Vieregg \,\orcidlink{0000-0002-4528-9886}}
\affiliation{Kavli Institute for Cosmological Physics, University of Chicago, 5640 South Ellis Avenue, Chicago, IL, 60637, USA}
\affiliation{Enrico Fermi Institute, University of Chicago, 5640 South Ellis Avenue, Chicago, IL, 60637, USA}
\affiliation{Department of Physics, University of Chicago, 5640 South Ellis Avenue, Chicago, IL, 60637, USA}
\affiliation{Department of Astronomy and Astrophysics, University of Chicago, 5640 South Ellis Avenue, Chicago, IL, 60637, USA}
\author{A.~Vitrier}
\affiliation{Sorbonne Universit\'e, CNRS, UMR 7095, Institut d'Astrophysique de Paris, 98 bis bd Arago, 75014 Paris, France}
\author{Y.~Wan}
\affiliation{Department of Astronomy, University of Illinois Urbana-Champaign, 1002 West Green Street, Urbana, IL, 61801, USA}
\affiliation{Center for AstroPhysical Surveys, National Center for Supercomputing Applications, Urbana, IL, 61801, USA}
\author{N.~Whitehorn\,\orcidlink{0000-0002-3157-0407}}
\affiliation{Department of Physics and Astronomy, Michigan State University, East Lansing, MI 48824, USA}
\author{W.~L.~K.~Wu\,\orcidlink{0000-0001-5411-6920}}
\affiliation{Kavli Institute for Particle Astrophysics and Cosmology, Stanford University, 452 Lomita Mall, Stanford, CA, 94305, USA}
\affiliation{SLAC National Accelerator Laboratory, 2575 Sand Hill Road, Menlo Park, CA, 94025, USA}
\author{M.~R.~Young}
\affiliation{Fermi National Accelerator Laboratory, MS209, P.O. Box 500, Batavia, IL, 60510, USA}
\affiliation{Kavli Institute for Cosmological Physics, University of Chicago, 5640 South Ellis Avenue, Chicago, IL, 60637, USA}
\collaboration{SPT-3G Collaboration}
\noaffiliation

\date{\today}
 
\begin{abstract}
We present a measurement of the $B$-mode polarization power spectrum of the cosmic microwave background anisotropies at 32 $\le$ $\ell$ $<$ 502 for three bands centered at 95, 150, and 220\,GHz using data from the SPT-3G receiver on the South Pole Telescope.  This work uses SPT-3G observations from the 2019 and 2020 winter observing seasons of a $\sim$1500\,deg$^2$ patch of sky that directly overlaps with fields observed with the BICEP/Keck family of telescopes, and covers part of the proposed Simons Observatory and CMB-S4 deep fields.  Employing new techniques for mitigating polarized atmospheric noise, the SPT-3G data demonstrates a white noise level of 9.3 (6.7) $\mu$K-arcmin at $\ell \sim 500$ for the 95\,GHz (150\,GHz) data,  with a $1/\ell$ noise knee at $\ell=128$ (182). 
We fit the observed six auto- and cross-frequency $B$-mode power spectra to a model including lensed $\Lambda$CDM $B$-modes and a combination of Galactic and extragalactic foregrounds. This work characterizes foregrounds in the vicinity of the BICEP/Keck survey area, finding foreground power consistent with that reported by the BICEP/Keck collaboration within the same region, and a factor of $\sim$ 3 higher power over the full SPT-3G survey area. Using SPT-3G data over the BICEP/Keck survey area, we place a 95\% upper limit on the tensor-to-scalar ratio of $r < \rlimitbksimple{}$ and find the statistical uncertainty on $r$ to be $\sigma(r) = \rsigmabksimple{}$. 
\end{abstract}
\maketitle
  
\section{Introduction}\label{sec:intro}
The $\Lambda$CDM cosmological model provides a precise framework for describing the evolution and composition of the universe. However, $\Lambda$CDM does not address fundamental questions about the initial conditions of the universe, such as the observed temperature uniformity of the cosmic microwave background (CMB) or the origin of large-scale structure.

One of the most compelling proposed extensions to $\Lambda$CDM is the theory of inflation \citep[e.g.,][]{guth81, kamionkowski2016quest}, which postulates a period of accelerating expansion in the very early universe. This rapid expansion stretches primordial quantum fluctuations to create tiny density fluctuations  that grow into the CMB anisotropies and eventually become the galaxies and stars that make up the large-scale structure of the universe today.  Inflation makes clear predictions for the initial conditions and geometry of the Universe such as flatness and a nearly scale-invariant power spectrum of primordial scalar perturbations, which have since been observationally confirmed \citep[e.g.,][]{planck18-6}.
Inflationary models also predict primordial tensor perturbations (gravitational waves), with an amplitude that depends on the energy scale of inflation and the shape of the inflationary potential.
The inflationary gravitational wave background is yet to be observed, and detecting these gravitational waves would provide new clues to the physics of the early universe.

The current best way to detect the signature of inflation is through the polarization patterns of the CMB. \citep{seljak1997signature} \citep{kamionkowski1997probe} CMB polarization can be decomposed into even-parity $E$~modes and odd-parity $B$~modes. Tensor perturbations from inflation are predicted to generate both  $E$~modes and $B$~modes. Scalar perturbations (density fluctuations) produce high levels of $E$-mode power but negligible $B$-mode polarization. This makes $B$-mode polarization a higher signal-to-noise observational channel for detecting the faint imprint of inflationary gravitational waves. However, Galactic emission also contributes to $B$-mode power at the large angular scales where the inflationary contributions are expected to peak ($\ell \approx$ 80). This power can be reduced by leveraging the different spectral behavior of Galactic foreground emission. 
Additionally, gravitational lensing mixes some $E$-mode power into $B$~modes; the technique of “delensing” \citep[e.g.,][]{kesden2002separation, ade2021demonstration,delmanzotti2017cmb} can be applied to reduce the lensing $B$~modes.

The power in gravitational waves can be parameterized in terms of the tensor-to-scalar ratio, $r$. Upper limits on $r$ from CMB $B$~modes have been set by experiments such as ABS \citep{kusaka18}, BICEP/Keck \citep{bicepkeck21c}, \textsl{Planck} \citep{tristram22}, \textsc{Polarbear} \citep{adachi22}, \textsc{Spider} \citep{spider22}, and SPTpol \citep{sayre20}, though no detection has been made. The most stringent constraints on inflationary gravitational waves comes from analyses of the most recent BICEP/Keck, WMAP, and \textsl{Planck} data releases \citep{bicepkeck21c, tristram22}, with the analysis from the BICEP collaboration reporting an upper limit of $r <$ 0.036.

This work uses data from the 2019 and 2020 austral winter seasons of the SPT-3G survey on the South Pole Telescope, covering $\sim$1500\,deg$^2$ centered at right ascension $0^\mathrm{h}$ and declination $-56^\circ$. The main survey field for SPT-3G was originally defined using iso-weight contours from the BICEP3 telescope \citep{bicep3}, as a proxy for the eventual footprint of the BICEP Array \citep{hui2018biceparray}. The SPT-3G region thus defined is centered on the BICEP/Keck field, but expands the sky area by a factor of 3.7 compared to the BICEP/Keck mask from the most recent $B$-mode analysis from the BICEP/Keck collaboration \citep{bicepkeck21c}, henceforth referred to as BK18. Additional sky coverage near the BICEP/Keck field is useful for characterizing foregrounds, the mitigation of which is essential for future surveys that cover a larger patch of sky than BICEP/Keck, such as those planned by the Simons Observatory and CMB-S4.

We present measurements of the $B$-mode auto- and cross-power spectra of the six possible combinations of SPT-3G maps at 95, 150, and 220\,GHz, employing an unbiased method of polarized atmospheric noise mitigation that uses the multi-chroic design of the SPT-3G detectors. We compare the measured bandpowers to a model that includes $\Lambda$CDM lensing-induced $B$~modes, Galactic dust, extragalactic radio galaxies, and inflationary gravitational waves to constrain both the tensor-to-scalar ratio $r$ and the properties of foreground emission over the observed sky region. 
 
This paper is organized as follows: Section~\ref{section:instrument} describes the SPT-3G instrument and observation strategy. Sections~\ref{section:tod} and~\ref{section:maps} describe the processing of raw time-ordered data into individual frequency maps, with a focus on the algorithm we use to mitigate contamination from polarized atmospheric emission and the apodization masks we use in this analysis. Section~\ref{section:powerspectrum} describes the process of turning these maps into measurements of the on-sky power spectrum and describes the steps and internal consistency checks we perform to ensure that our power spectra are not significantly biased by instrumental systematics. We describe the model and likelihood in Section~\ref{section:model}. We present the measured CMB power spectra and constraints on $r$ and foreground emission in Section~\ref{section:results} and conclude in
Section~\ref{section:conclusions}.

\section{Instrument and Observations}\label{section:instrument} The South Pole Telescope \citep[SPT,][]{carlstrom11} is a 10-meter-diameter off-axis Gregorian telescope designed for observations of the CMB. The SPT-3G receiver focuses incident light from the sky onto an array of broadband dual-polarization antennas. The signal from each antenna is split into three bands centered at 95, 150, and 220\,GHz via in-line band-defining filters. These couple to approximately 16,000 transition-edge sensor bolometers. This work utilizes data from the SPT-3G receiver acquired during the 2019 and 2020 austral winter observing seasons. The data cover a  $\sim$1500\,deg$^2$ area centered at right ascension 0h and declination $-56^{\circ}$. Observations are divided into four overlapping raster-scanned subfields, each of which is the full width of the field (from $20^\mathrm{h}40^\mathrm{m}$ to $3^\mathrm{h}20^\mathrm{m}$) in right ascension and roughly one quarter ($7.5^{\circ}$) of the field height in declination.  Observations consist of constant-elevation scans in azimuth, with steps in elevation after one right-going and one left-going scan. With an azimuthal scanning rate of 1 degree per second, each scan lasts approximately 100 seconds, and a single subfield observation is completed in $\sim$2.5 hours. The temperature response of the instrument is calibrated with regular observations of H\textsc{II} regions of known flux, specifically MAT\,5a and RCW\,38. More details on the instrument and calibration strategy can be found in \citet{sobrin22}.
 
\section{Data Reduction}\label{section:tod}
Following previous SPT analyses \citep[e.g.,][]{henning18, sayre20, dutcher21}, we use an observation-by-observation filter-and-bin mapmaker. This approach is advantageous due to its simplicity and the ability to perform observation-by-observation map processing steps to match changes in noise, including those due to polarized atmospheric emission.
First, a series of data cuts removes low-quality data from the processing pipeline. Exclusion reasons include an excess of ``glitches'' (statistically unlikely deviations from a rolling average), lack of calibration information, improperly biased detectors, low signal-to-noise on calibration sources, or readout errors. We analyze the remaining data on a scan-by-scan basis. We low-pass filter each scan of time-ordered-data (TOD) with a filter cutoff of $\ell = 3000$ to reduce aliasing of high-frequency noise beyond the spatial Nyquist frequency of the map pixel size. We high-pass filter the data by subtracting a 10th-order polynomial to reduce low-frequency noise on spatial scales larger than $\sim$10$^{\circ}$.

In contrast with previous SPT power spectrum analyses, two additional operations are performed to specifically target low-$\ell$ polarized noise before the individual detector data are averaged into a map. First, to reduce temperature-to-polarization leakage, we calibrate the relative detector gains in each frequency band for each pair of orthogonal polarizations in a pixel. The gain factors are determined by minimizing the difference in intra-pixel response to unpolarized atmospheric signal during the telescope elevation slews that precede each field observation. This recalibration mitigates gain mismatches within pixel pairs that arise from variations in frequency bandpasses or beam profiles between detectors. This additional calibration step offers a $\sim$20\% improvement in polarized large-angular-scale noise ($\ell < 150$) over the typical SPT method of calibrating relative detector gains using astrophysical sources. It is interesting to note that, unlike previous SPT-3G analyses, we detect negligible ($\lesssim0.1\%$) monopole temperature-to-polarization leakage below $\ell = 500$ in the final maps.

\begin{figure*}[htbp]
    \begin{center}
        \includegraphics[width=\linewidth]{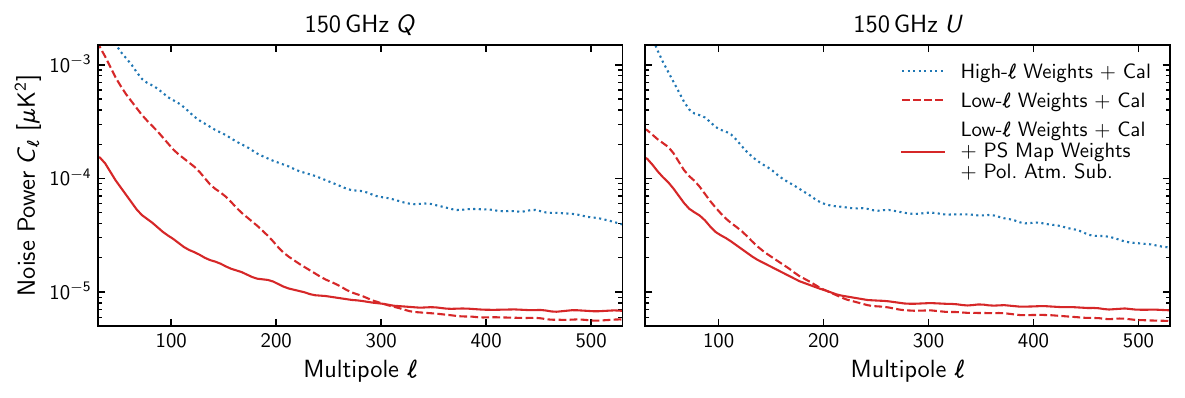}
        \caption{Noise curves using different atmospheric contamination mitigation techniques. \textit{Left:} $Q$ 150\,GHz noise power, \textit{Right:} $U$ 150\,GHz noise power. The dotted blue curve in each panel is the noise level using weighting and calibration choices typically used for SPT-3G maps optimized for high-$\ell$ science, namely H\textsc{II} region gains and TOD weighted by the 1-4 Hz white noise region in temperature. It should be noted that for such analyses a much stronger high-pass filter is applied, which results in a complete loss of low-$\ell$ information but a significantly reduced white noise level (below the red curves). The dashed red curve in each panel shows the improvement from implementing polarization-sensitive low-frequency weights, and determining relative detector gains from elevation slews. These improvements help more in $U$ than in $Q$. The residual $Q$/$U$ asymmetry highlights the impact of polarized atmosphere, which can be directly targeted by power-spectrum-based map weights and polarized atmosphere map-based subtraction (solid red line; see Section~\ref{sec:polatm}, \cite{coerver24}). Red curves adapted from \cite{coerver24}.}
        \label{fig:noisecurves}
    \end{center}
    \vspace{-15pt}
\end{figure*}

Additionally, to further reduce polarized noise with relative detector weights, we form ``pair-difference'' TOD by differencing the detector TOD within the same pixel. This ``pair-difference'' TOD subtracts the common component of the dominant signal in each detector's TOD, the unpolarized atmospheric signal, allowing for the variance to be dominated by polarized noise. We then give both intra-pixel detectors the same weight, assigned from the inverse variance of low-frequency (0.1–1\,Hz) noise power in the pair-difference TOD. This weighting approach downweights detectors with elevated non-common low-frequency polarized noise. 
This approach yields a substantial improvement in low-$\ell$ polarized noise performance, as illustrated by the transition from the blue to red dashed curves in Figure~\ref{fig:noisecurves}. The blue curves reflect the noise levels using weighting and calibration choices typically used for SPT-3G maps optimized for high-$\ell$ temperature analyses, whereas the red dashed curves show the improvement from implementing pair-difference-based low-frequency weighting and gain calibration using elevation slews. These methods significantly reduce large-scale polarized noise, particularly in $U$, by targeting polarized noise sources prior to mapmaking.

\section{Maps}\label{section:maps}
We bin the filtered and weighted detector TOD into Stokes $T$, $Q$, and $U$ HEALPix \citep{gorski05} maps on an observation-by-observation basis, using methods established in previous SPT analyses \citep[e.g.,][]{dutcher21}. Before combining these single-observation maps into a final, full-depth, weighted average of all the observation maps, or ``coadd,'' we implement two additional processing steps to mitigate contamination from polarized atmospheric emission.

\subsection{Mitigation of Polarized Atmosphere}\label{sec:polatm}
We find polarized atmosphere to be the largest source of polarized noise in SPT-3G large-angular-scale ($\ell \lesssim 200$) data. This polarized emission, primarily arising from horizontally aligned ice crystals in the atmosphere, manifests as excess noise in the horizon-aligned Stokes $Q$ polarization on large angular scales. The spatial distribution of these ice crystals follows the spatial scaling expected from Kolmogorov turbulence, and their polarized emission shows a steep frequency dependence consistent with a combination of Rayleigh scattering and thermal emission. More details on the characterization of polarized atmosphere at the South Pole and the mitigation strategies we describe can be found in \citet{coerver24}.

To mitigate polarized atmospheric noise in the 150\,GHz data, we subtract a scaled version of the 220\,GHz $Q$ map from each 150\,GHz $Q$ map on an observation-by-observation basis. Prior to this subtraction, the full-depth coadd (a high signal-to-noise map of fixed sky signals) is removed from each 220\,GHz observation map, and a Butterworth low-pass filter is applied at the angular scale where the map noise becomes dominated by detector noise. These steps isolate the changing atmospheric component in the 220\,GHz map, which is then used to clean the 150\,GHz map. This technique utilizes the stable frequency scaling of the atmospheric polarization and our co-pointing tri-chroic detectors.  
This improves noise levels for  $\sim$70\% of observations, reducing large-angular-scale noise without affecting the CMB and other fixed sky signals or introducing bias. Such a subtraction is not necessary for the 95\,GHz data since the atmospheric spectrum is so steep with frequency. 
 
After cleaning each observation map, we create the full-depth coadd and data subset, or ``bundle,'' coadds with a weighted average of observations.  These weights are determined from the low-frequency noise ($\ell = 50-250$) in the $Q$ and $U$ power spectra of each observation map. This downweights maps with noise correlated between detectors such as that expected from the polarized atmosphere. This weighting ensures that lower-noise observations contribute more to the final data products. Together, these methods effectively mitigate atmospheric contamination, leading to a 5$\times$ improvement in polarized noise power in the range $\ell \sim 30-100$, the regime important for sensitivity to inflationary $B$~modes. The impact of both the map-based and TOD-based mitigation techniques is shown in Figure~\ref{fig:noisecurves} and Table~\ref{tab:ellknees}. The transition from the red dashed to solid red curve in Figure~\ref{fig:noisecurves} highlights the additional benefit of incorporating power-spectrum-derived observation weights and direct subtraction of polarized atmospheric emission using the 220\,GHz maps. These map-based optimizations provide a further reduction in large-scale polarized noise, particularly in $Q$, and are useful tools for achieving the low noise levels necessary for robust $B$-mode sensitivity at $\ell <200$.

\begin{table}[t]
\centering
\setlength{\tabcolsep}{6pt} 
\renewcommand{\arraystretch}{1.2}
\begin{tabular}{c|cc|cc}
\toprule
Map-based & & & \multicolumn{2}{c}{Noise [$\mu$K-arcmin]} \\
optimization & Band & $\ell_\mathrm{knee}$ & ~high-$\ell$ & $\ell = 80$ \\
\midrule
no & 95\,GHz & 166 & ~8.6 & 23.1 \\
& 150\,GHz & 259 & ~6.7 & 43.9 \\
& 220\,GHz & 318 & ~26.5 & 247.7 \\ \midrule
yes & 95\,GHz  & 128 & ~9.3 & 18.0 \\
& 150\,GHz & 182 & ~7.6 & 19.7 \\
& 220\,GHz & 144 & ~32.2 & 68.1 \\
\bottomrule
\end{tabular}
\caption{Table of the $\ell_\mathrm{knee}$ and white (high-$\ell$) noise level in $\mu$K-arcmin for different polarized atmosphere optimization strategies. The data without map-based optimization include 10th order polynomial filtering, 0.1--1\,Hz pair-difference polarized weights, and elevation slew gains as described in the text (the red dashed curve in Figure~\ref{fig:noisecurves}). The data with map-based optimization include the same steps as above, in addition to power spectrum map weights and map-based polarized atmosphere subtraction for the 150\,GHz data as described in the text (the solid red curve in Figure~\ref{fig:noisecurves}). The $\ell_\mathrm{knee}$ is reduced with the map-based renormalization steps, but the white noise amplitudes increase due to the reduction in effective data volume from weighting.}
\label{tab:ellknees}
\end{table}

\subsection{\texorpdfstring{$QU$}{QU} Rotation}\label{section:qu} 

The linear polarization angle of each individual detector is set by its orientation in the focal plane. After installation, the nominal angles are checked with observations of Centaurus A (see \citealt{sobrin22} for details). Residual errors in these angles result in
two effects: 1) random detector-to-detector variations result in loss of polarization efficiency, which is folded into the overall polarization calibration factor (Section~\ref{sec:cal}); 2) an error in the mean angle results in an effective rotation of the polarization 
coordinate system, which in turn results in the following relationship between observed and true $Q$ and $U$ skies:
\begin{equation}
\tilde{Q} + i \tilde{U} = \left(Q+iU\right)e^{2i\Delta\psi},   
\end{equation}
where $\Delta\psi$ is the mean angle error. To correct for this, we perform the global angle rotation of the $Q$ and $U$ maps that minimizes the $TB$ and $EB$ spectra, following the expectations in \citet{seljak1997signature,kamionkowski1997probe}.

We rotate the 95\,GHz maps by $\Delta\psi = 0.008$ radians, the 150\,GHz maps by $\Delta\psi = 0.007$ radians, and the 220\,GHz maps by $\Delta\psi = -0.006$ radians.

\subsection{Apodization Mask and Maps}\label{section:apod}

\begin{figure}[t]
    \begin{center}
        \includegraphics[width=1.0\linewidth, keepaspectratio]{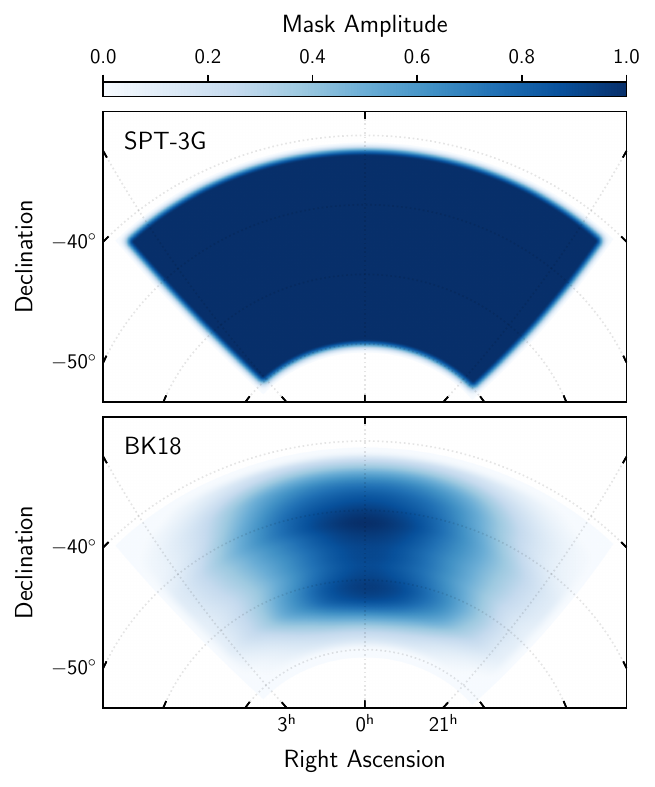}
        \caption{The two apodization masks used in this work. \textit{Top:} SPT-3G 1500\,deg$^2$ field ($f_{\mathrm{sky}}$ = 0.037). \textit{Bottom:} BK18 mask ($f_{\mathrm{sky}}$ = 0.010).}
        \label{fig:apod}
    \end{center}
\end{figure}

As discussed in Section~\ref{sec:intro}, the SPT-3G survey region is centered on the BICEP/Keck field but extends beyond its core area. The sky coverage difference can be seen in the top and bottom panels of Figure~\ref{fig:apod}. There are two reasons for the wider area. First, the SPT-3G region was designed to match the larger footprint of the BICEP3 and future BICEP Array fields, which partially accounts for the larger sky coverage of SPT-3G vs.~BK18. Second, the SPT-3G field-of-view is smaller than that of the BICEP Array, which leads to a sharper transition from zero weight areas outside the survey to the complete coverage region of the survey.
\begin{figure*}[t]
    \begin{center}
        \includegraphics[width=\linewidth]{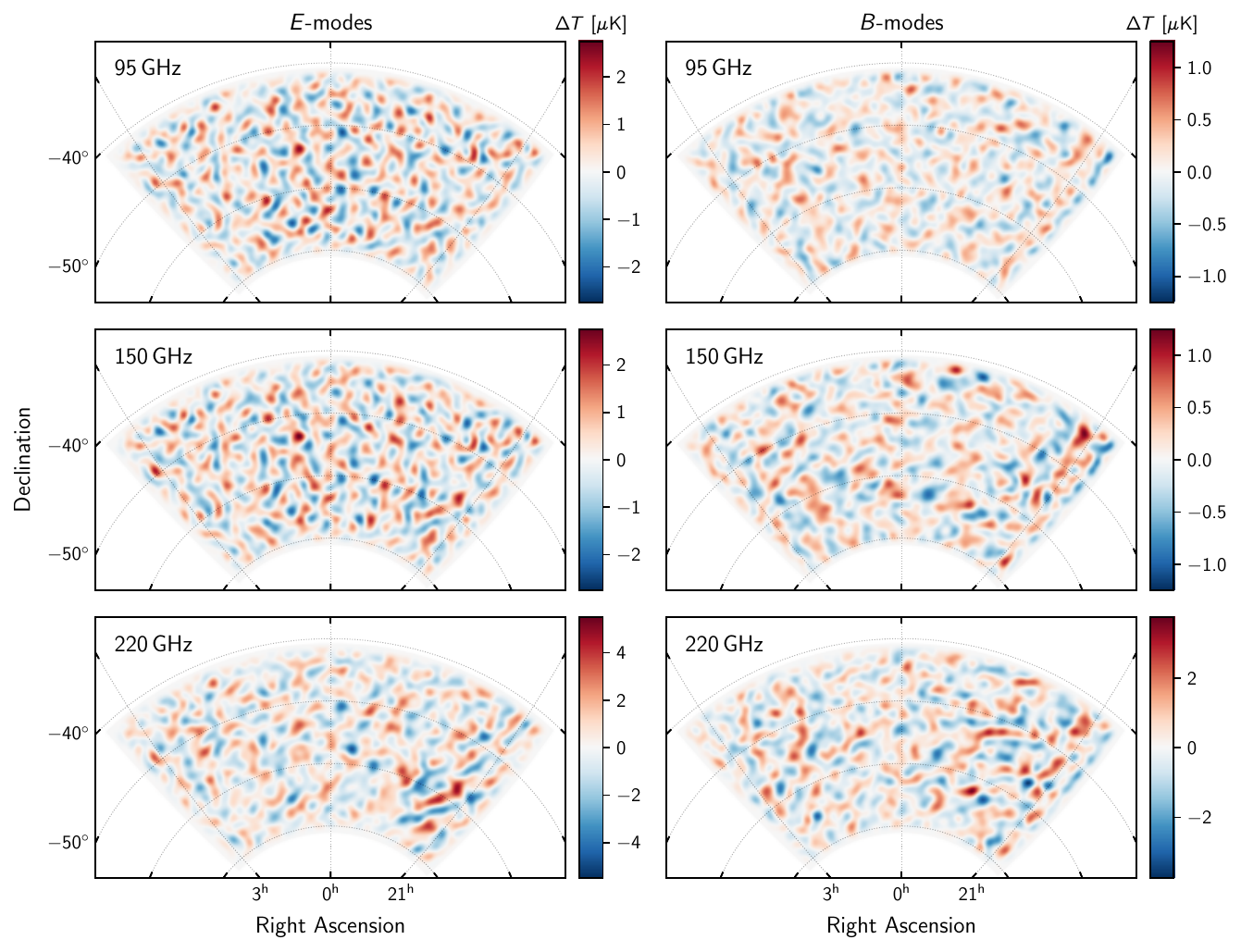}
        \caption{$E$-mode and $B$-mode maps for the 1500\,deg$^2$ SPT-3G field. A bandpass filter over $\ell \in 30-150$ has been applied to visually highlight the large angular scales where the inflationary signal peaks. Elevated levels of foreground emission can be seen on the right side of the map, especially at 220\,GHz, noting that the BK18 mask in Figure~\ref{fig:apod} avoids these regions.}
        \label{fig:Emaps}
    \end{center}
    \vspace{-15pt}
\end{figure*}

For the apodization mask for the SPT-3G field, we first create a binary mask to avoid areas with low SPT-3G data weight. We then calculate the distance to the edge $\theta$ and multiply the respective binary masks by the functional form $0.5 \left(1 - \cos\left(\pi \cdot \theta / 2^\circ \right)\right).$ This smoothly adjusts the mask to zero within $2^\circ$ of the edges, mitigating edge effects in the analysis. 
We upsample and smooth the published BK18 mask to accommodate the higher pixel resolution. 
The SPT-3G 1500\,deg$^2$ apodization mask and the upsampled and smoothed BK18 apodization mask can be seen in Figure~\ref{fig:apod}. 

The final SPT-3G $E$-mode and $B$-mode maps for the 1500\,deg$^2$ field are presented in Figure~\ref{fig:Emaps}. The higher levels of foregrounds in the SPT-3G field can be seen in the bottom right-hand corner, motivating the choice of the BK18 mask for the baseline science result.

\subsection{Noise realizations}
To get an accurate estimate of the $B$-mode power spectrum noise, we generate ``signflip'' noise maps by assigning a random +1 or $-1$ sign to each observation map and combining them in a weighted average. To remove residual astrophysical signal resulting from non-uniform weights, we construct a template of the sky from the coadd of all observations and subtract that signal from each of the observation maps before they are combined to produce the signflip noise estimate. This subtraction ensures that the astrophysical signal is effectively canceled, leaving a realization of the observational noise.  Because of our large number of observations, this method provides an estimate of the noise without the necessity of a complete instrument noise model. Noise spectra are calculated by averaging the spectra of 100 of these signflip maps.
 
\section{Power Spectrum Methods}\label{section:powerspectrum} We use the NaMaster library \citep{alonso19} to calculate pure $B$-mode maps and power spectra from the measured $Q$ and $U$ maps. We calculate all auto- and cross-power spectra between our 3 bands, leading to a total of 6 spectra: $95 \times 95$\,GHz, $95 \times 150$\,GHz, $95 \times 220$\,GHz, $150 \times 150$\,GHz, $150 \times 220$\,GHz, and $220 \times 220$\,GHz. 
 
The measured spectra are related to the true sky power by \citet{hivon02} 
\begin{equation}
    \langle\hat{C}_{\ell}\rangle =  F_{{\ell}} B_{{\ell}}^2 \langle C_{{\ell}}\rangle,
\end{equation}
where $\langle\hat{C}_{\ell}\rangle$ is the ensemble average of the biased power spectrum estimate from NaMaster and $\langle C_{{\ell}}\rangle$ is the ensemble average of the true sky power spectrum. 
To measure $\langle\hat{C}_{\ell}\rangle$, we use an average over the cross-bundle power spectra from NaMaster of 24 chronological equal-weight data bundles. The use of bundle cross-spectra is performed to eliminate noise bias. The NaMaster estimate of $\langle\hat{C}_{\ell}\rangle$ has been corrected for cross-coupling of power between different $\ell$ modes \citep{alonso19}. $F_{{\ell}}$ is the effective filter transfer function induced by the time-domain filtering of the TOD and pixelation of the map, while the beam, $B_{{\ell}}^2$, describes the instrument response to a point source.

We calculate $F_{{\ell}}$ by creating mock TOD from 100 simulated CMB skies. Each sky is a Gaussian random realization of ${C}_{\ell}$ that include $\Lambda$CDM CMB with $r$ = 0 plus Galactic dust with levels matched to the region of the sky following the model described in Section~\ref{section:model}. We then process these TOD through our filtering, mapmaking and power spectrum estimation pipeline, and calculate the ratio with respect to the input ${C}_{\ell}$. 

The beam, $B_{{\ell}}^2$, is measured using a combination of dedicated Saturn observations and bright sources in the SPT-3G survey field. The Saturn observations provide a high signal-to-noise ratio on the large-scale beam side lobes ($\sim$1 degree). However, Saturn is so bright that it saturates some of the detectors when they pass directly over the disk of the planet. Therefore, we excise the main lobe of the beam from the Saturn maps (pixels at a distance $\lesssim 1\farcm5$ from the center of Saturn), and stitch in maps of the bright sources in the survey field to recover the main lobe. We perform this stitching operation on all possible pairs of Saturn and bright source maps, then construct cross-spectra of the stitched maps to calculate $B_\ell$. As reported in \citet{ge2024cosmology}, there is evidence the sidelobe level for the polarized beam is lower than measured for the intensity beam. We performed our analysis both assuming the polarized beam is identical to the intensity beam and again with the depolarized sidelobe model and find consistent results. The results presented in the following sections use the intensity beam. Following \citet{hivon02} we bin our final bandpowers using weights $w$ corresponding to 
 \begin{equation}
w = \frac{\left(C_\ell^{BB_{\mathrm{IGW\,theory}}}\right)^2}{\mathrm{var}\left(C^{BB}_\ell\right)},
 \end{equation}
where the numerator is an inflationary gravitational wave (IGW) theory curve and the denominator is the variance in the $B$-mode power spectrum calculated from the diagonal of the covariance matrix. Therefore we weight our final bandpowers on signal-to-noise, with bins of $\ell = 32-102$, $\ell = 102-202$, $\ell = 202-302$, $\ell = 302-402$, and $\ell = 402-502$.

\subsection{Power Spectrum Calibration Terms}\label{sec:cal}
We apply two additional calibration terms to the final power spectra: absolute polarization calibration and $EB$/$TB$ subtraction.

 We estimate the absolute polarization calibration in each frequency band by comparing the measured $EE$ power spectra at $302 \le \ell < 602$ to the best-fit $\Lambda$CDM model for the \textsl{Planck} 2018 \texttt{base\_plikHM\_TTTEEE\_lowl\_lowE\_lensing} data combination. 
 We assume uncertainties in the \textsl{Planck} model spectrum are negligible compared to our measurement uncertainty. 
 We find  calibration uncertainties of $2.5\%,~ 2.3\%, $ and $ ~3.6\%$ respectively for 95, 150, and 220\,GHz. 
 
Following \citet{sayre20}, we subtract monopole $E$ to $B$ and $T$ to $B$ leakage contributions to the observed spectra $D_b^{'BB}$ to recover the true spectra $D_b^{BB}$ via
 \begin{equation}
D_b^{'BB} = D_b^{BB} + \frac{\left(D_b^{EB}\right)^2}{D_b^{EE}}  + \frac{\left(D_b^{TB}\right)^2}{D_b^{TT}}   
 \end{equation}
where $D_l \equiv \ell(\ell+1)C_l / 2\pi$ and $b$ denotes the bin.
This subtraction compensates for any residual leakage after the $QU$ rotation described in Section~\ref{section:qu}. The second term accounts for correlations between $B$ and $E$ from, for example, miscalibration of polarization angles, beam effects, or $E$ to $B$ filtering leakage. The third term accounts for correlations between $B$ and $T$ from an error in the absolute polarization angle or $T$ to $B$ filtering leakage, among other effects. In the lowest $\ell$ bin, these terms are less than $3 \times 10^{-4} \mu \textup{K}^2$ for $95 \times 95$\,GHz, less than $6 \times 10^{-4} \mu \textup{K}^2$ for $150 \times 150$\,GHz, and less than $2 \times 10^{-3} \mu \textup{K}^2$ for $220 \times 220$\,GHz.

Unlike $B$-mode experiments with small aperture telescopes with degree-scale beam sizes, SPT-3G has a small arcmin-scale beam. Therefore, we can neglect contributions to the power spectra from higher-order beam leakage. For example, for dipole leakage, the strength of the leakage depends on the dipole moment, which comes from systematic pointing offsets correlated with polarization angle. For us to have a large enough dipole moment to bias the angular scales where inflationary $B$-modes peak, our pointing offsets would have to be wrong by a full beam width. We know the offsets are much smaller from the angular size of the point sources in our maps. The same argument follows for higher-order leakage terms.
 
\subsection{Null Tests}\label{sec:nulls} To look for evidence of systematic contamination of the data, we carry out the following ``null tests'' following the methods of \citet{dutcher21} and previous SPT analyses.  We split the data in five ways: 1) chronologically, 2) based on whether the sun is above or below the horizon, 3) based on whether the moon is above or below the horizon, 4) the azimuth of the field, and 5) the direction of motion of the telescope.  For each null test, we construct a ``null spectrum'' by dividing the maps into two bundles depending on the systematic effect in question, then subtract the bundles to null any sky-locked contributions. This leaves a spectrum that is maximally sensitive to the systematic effect. We then compare the null spectrum to an ``expectation spectrum'' constructed by multiplying individual observation maps in both bundles by +1 or $-1$, then subtracting the bundles to provide an estimate of the noise bias. The expectation spectrum is then computed 100 times for 100 different random +1/$-1$ combinations. Then we calculate the $\chi^2$ of the null spectrum compared to the expectation spectrum. From this, we calculate the probability to exceed (PTE) this $\chi^2$ value, given the number of degrees of freedom, using the cumulative distribution function. The PTE values of the different bands and null tests can be seen in Table~\ref{tab:nulls}.  

Following \cite{dutcher21}, we set passing criteria for the null tests as: 1) The entire table of PTE values is consistent with a uniform distribution between 0 and 1, with a Kolmogorov-Smirnov (KS) test p-value $>$ 0.05; 2) individual PTE values are larger than $0.05 / N_{\mathrm{tests}} = 0.0008$, where $N_{\mathrm{tests}}$ is the number of bands $\times$ number of null tests $\times$ number of polarization combinations; 3) using Fisher's method, the combination of PTEs across each band for $BB$/$EB$/$TB$/$EE$, for each test, has a PTE above $0.05 / N_{\mathrm{rows}} = 0.01$ where $N_{\mathrm{rows}}$ is the number of null tests. All of these tests pass: the data have a KS test p-value of 0.61, all individual PTE values are larger than 0.0008 as shown in Table~\ref{tab:nulls}, and the Fisher PTE values all exceed 0.01 with values of 0.82 (azimuth), 0.29 (chron), 0.86 (moon), 0.18 (scan dir), and 0.38 (sun). 

\begin{table}[t]
\setlength{\tabcolsep}{6pt}
\centering
\renewcommand{\arraystretch}{1.2}
\begin{tabular}{cc|cccc}
\toprule
Band & Null Test  &  $BB$ &  $EB$ &  $TB$  & $EE$ \\
\midrule
95\,GHz & azimuth        &   0.4590    &   0.8989  &  0.1645  &  0.3783   \\
& chron     &   0.3643   &   0.1007  &  0.6008  &  0.0711  \\
& moon      &   0.6677   &   0.4731  &  0.9950   &  0.1028   \\
& scan dir   &   0.0564  &   0.3512  &  0.1047  &  0.5628   \\
& sun       &   0.1326   &   0.4059  &  0.4908  &  0.0227  \\
\midrule
150\,GHz & azimuth         &   0.9057  &   0.5470   &  0.2964  &  0.9920   \\
& chron      &   0.5088  &   0.6891  &  0.8624  &  0.5716  \\
& moon       &   0.6651  &   0.7043  &  0.6330   &  0.3171  \\
& scan dir    &   0.6737  &   0.3872  &  0.6984  &  0.9002  \\
& sun        &   0.4722  &   0.5360   &  0.7980   &  0.7508  \\
\midrule
220\,GHz & azimuth         &   0.5218   &   0.1227  &  0.8817   &  0.6889   \\
& chron      &   0.3816   &   0.5205  &  0.0972  &  0.2228   \\
& moon       &   0.4158   &   0.6601  &  0.3119   &  0.8766   \\
& scan dir    &   0.0577  &   0.3904  &  0.9461   &  0.0746  \\
& sun        &   0.3521   &   0.2457  &  0.8367   &  0.4394  \\
\bottomrule
\end{tabular}
\caption{Individual null test PTE values for each test across all frequencies and spectra. The individual PTE values are all larger than 0.0008 and their statistics in ensemble pass the tests described in Section~\ref{sec:nulls}. We find no evidence for systematic biases.}
\label{tab:nulls}
\end{table}

\subsection{Covariance}\label{sec:Covariance}The covariance matrix encapsulates both the variance and bin-to-bin correlations of the bandpowers, including between frequency bands. This is necessary to construct the likelihood and for cosmological parameter constraints.
 
We use 100 signal-plus-noise simulations to calculate the covariance matrix.  For each mask, we generate a set of 100 signal realizations as described in Section~\ref{section:powerspectrum}, with a level of Galactic dust power matched to the dust level within the mask footprint. Each signal-plus-noise simulation is then constructed by summing one signal realization with one randomly chosen signflip noise realization. This approach ensures that the covariance captures the relationship between signal and noise contributions while accounting for the actual characteristics of the instrumental noise. Since this is a noisy realization of the true covariance matrix, we ``condition'' the estimate by zeroing off-diagonal elements that are statistically consistent with noise, ensuring the matrix better reflects the true underlying correlations.

\begin{table*}
\setlength{\tabcolsep}{6pt}
\centering
\renewcommand{\arraystretch}{1.2}
\begin{tabular}{c|cccccc}
\toprule
{$\ell$ Center} & {$D_{\ell}$ 95x95} & {$D_{\ell}$ 95x150} & {$D_{\ell}$ 95x220} & {$D_{\ell}$ 150x150} & {$D_{\ell}$ 150x220} & {$D_{\ell}$ 220x220} \\
\midrule
75  & 0.009 (0.005) & 0.020 (0.006) & 0.006 (0.015) & 0.042 (0.009) & 0.053 (0.025) & 0.165 (0.113) \\
150 & 0.010 (0.006) & 0.001 (0.006) & 0.009 (0.017) & 0.012 (0.007) & 0.026 (0.024) & 0.211 (0.092) \\
245 & 0.039 (0.006) & 0.026 (0.004) & 0.028 (0.012) & 0.032 (0.006) & 0.055 (0.015) & 0.133 (0.067) \\
346 & 0.034 (0.009) & 0.033 (0.006) & 0.047 (0.024) & 0.040 (0.008) & 0.053 (0.023) & -0.058 (0.114) \\
447 & 0.057 (0.013) & 0.049 (0.008) & 0.061 (0.033) & 0.063 (0.011) & 0.084 (0.030) & 0.341 (0.138) \\
\bottomrule
\end{tabular}
\caption{Values of $D_{\ell}$ and $\sigma(D_{\ell})$ for each bandpower bin for the SPT-3G data with the BK18 apodization mask. Each cell lists the value and uncertainty in $\mu \mathrm{K}^2$.}
\label{tab:bandpower_error}
\end{table*}

\begin{figure*}[t]
    \begin{center}
        \includegraphics[width=\linewidth]{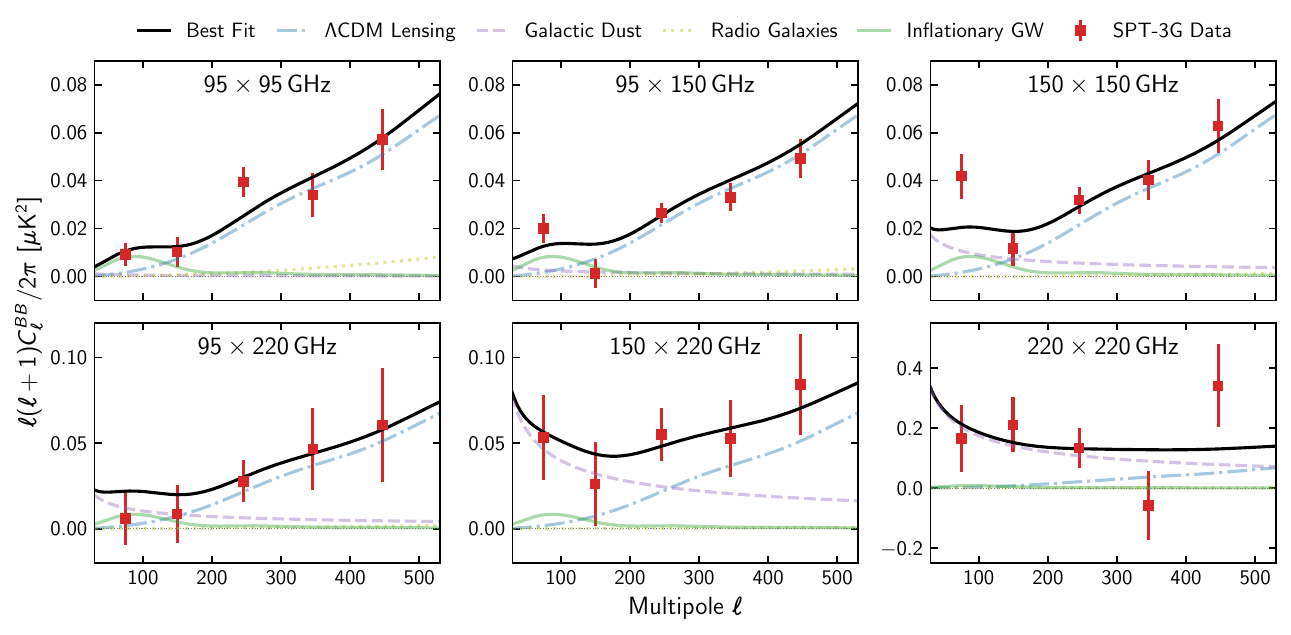}
        \caption{SPT-3G $B$-mode bandpowers for the BK18 mask for (top row) $95 \times 95$\,GHz, $95 \times 150$\,GHz, $150 \times 150$\,GHz, (bottom row) $95 \times 220$\,GHz, $150 \times 220$\,GHz, and $220 \times 220$\,GHz. The bandpowers with their 1$\sigma$ error bars calculated from the covariance matrix are presented in red. The black line shows the best-fit model from Section~\ref{section:results}. The colored lines show the best-fit model broken up by component. }
        \label{fig:bandpowers}
    \end{center}
    \vspace{-15pt}
\end{figure*}
 
\section{Modeling and Likelihood}\label{section:model}
To derive parameter constraints, we compare the measured bandpowers to modeled power spectra. 
We do this with the Hamimeche-Lewis (H\&L) likelihood approximation \citep{hamimeche08} to account for the expected non-Gaussian distribution of bandpowers. This method approximates the exact likelihood by applying a transformation to the covariance matrix to be used in the likelihood 
\begin{equation}
-2\ln\mathcal{L}\left(C_\ell|\hat{C_\ell}\right) = \mathbf{X}_{g\ell}^T \mathbf{M}_{f\ell}^{-1} \mathbf{X}_{g\ell}.
\end{equation}
$\mathbf{M}_{f\ell}$ is the fiducial model covariance block diagonal matrix calculated from the signal and noise sims referenced above, a fiducial $r=0$ model spectrum, and noise calculated from signflips. Using $\mathrm{vecp}$, the vector of distinct matrix elements, we construct
\begin{equation}
\left[\mathbf{X}_g\right]_\ell~\equiv~\mathrm{vecp}\left(C_{f\ell}^{1/2}g\left[C_{\ell}^{-1/2}\,\hat{C_\ell}\,C_{\ell}^{-1/2}\right]C_{f\ell}^{1/2}\right)
\end{equation}
where $g(x) \equiv \mathrm{sign}(x-1) \sqrt{2(x-ln(x)-1)}$, and $C_{\ell}$ is the model spectrum for a sampled set of parameters. 

All parameter constraints in this work are derived using Markov Chain Monte Carlo, as implemented in the Cobaya package \citep{torrado21}.
  
\subsection{Model}

We fit the observed $B$-mode power to a 8-component model that includes $r$, three calibration terms described in Section~\ref{sec:cal}, lensing $B$~modes (fixed to the expectation for a \textsl{Planck} $\Lambda$CDM cosmology), a Poisson term for radio galaxies, and Galactic dust emission. 
We report upper limits on $r$ with a uniform prior for $r>0$.
 
As in previous works (e.g. BK18), we assume the dust spectral energy distribution (SED) follows a modified black body spectrum, $\nu^\beta B_\nu(T)$, where $\nu$ is the observing frequency, $\beta$ the index of the power law, and $B_\nu(T)$ is a black body spectrum of temperature $T$.
We fix the temperature to $T=19.6$\,K, as in BK18 and \citet{akrami2020planckfg}.
We set a broad uniform prior on $\beta$ of $\beta\in [0,3]$. 
We model the angular dependence of the Galactic dust
with a simple power law
\be
D_\ell = A^{\mathrm{150\,GHz}}_{\ell=80} \left( \frac{\ell}{80}\right)^\alpha,
\ee
with two free parameters, $A^\mathrm{{150\,GHz}}_{\ell=80}$ the amplitude at 150\,GHz and $\ell=80$, and the power-law index, $\alpha$.
We use uniform priors on both parameters: $A^{\mathrm{150\,GHz}}_{\ell=80} > 0$ and $\alpha \in [-1.2,-0.2]$.

We model Poisson power from radio galaxies \citep{gupta2019fractional, thorne2017python} as 
\be
D_\ell = A^{\mathrm{95\,GHz,rg}}_{\ell=500} \left( \frac{\ell}{500}\right)^2. 
\ee
 We assume the spectral energy distribution $S_\nu$  of these galaxies follows $S_\nu \propto \nu^{-0.75}$. 
We set a  positive prior on the amplitude $A^{\mathrm{95\,GHz,~ rg}}_{\ell=500}$.

In previous analyses from the BICEP/Keck collaboration \citep{bicepkeck21c, array2018bicep2}, an additional term was included to model the synchrotron emission from the Galaxy.  The amount of synchrotron power from the BK18 fits would be a negligible contribution relative to the statistical uncertainty of our measurement.   To constrain synchrotron power outside the BK18 field, we extrapolate the polarized synchrotron power measured by WMAP data at 23\,GHz \citep{wmap1,wmap2} over the SPT-3G 1500\,deg$^2$ field to 95\,GHz. This predicts a synchrotron power of $1.5\times10^{-5}\, \mu \textup{K}^2$ in the first bin and even less in other bins. As this is a factor of ~300 times smaller than the bandpower uncertainty in the first bin, we choose to neglect synchrotron emission.

\section{Measured Bandpowers and Upper Limits on the Tensor-to-Scalar Ratio}\label{section:results}
We now present measured bandpowers from the SPT-3G data and resulting constraints on the tensor-to-scalar ratio. 
We first show in Section ~\ref{section:bkfit} the $B$-mode power spectrum measured in our baseline analysis using the SPT-3G data over the BK18 survey area, along with derived parameter constraints on the tensor-to-scalar ratio $r$ and foreground emission. 
We then look at the bandpowers measured across the full SPT-3G 1500\,deg$^2$ survey area in Section~\ref{section:fullfit}, and discuss how elevated foreground emission in this larger region impacts the recovered cosmological constraints. 

The bandpowers, covariance, and other data products are available on the SPT website\footnote[1]{https://pole.uchicago.edu/public/data/zebrowski25}, along with a likelihood implementation. 

\subsection{BK18 Survey Area}\label{section:bkfit}

\begin{table*}
\setlength{\tabcolsep}{6pt}
\centering
\renewcommand{\arraystretch}{1.2}
    \begin{tabular}{l|cc|cccc|c}
        \toprule
        Apodization Mask  &  95\% CL & $\sigma(r)$ & $A^{\mathrm{150\,GHz}}_{\ell=80} $ & $\alpha$ & $\beta$  &$A^{\mathrm{95\,GHz,\,rg}}_{\ell=500}$ &  PTE\\ \midrule
        BK18 &  $r<\rlimitbksimple{}$&\rsigmabksimple{}&\Deightyonefiftyghzconstraintbksimple{}& \alphagalconstraintbksimple{} & \betagalconstraintbksimple{}  & \Dfivehundredninetyfiveghzsyncpoissonconstraintbksimple{} 
        & 42\%\\
        SPT-3G 1500\,deg$^2$  &  $r<\rlimitfullsimple{}$ &\rsigmafullsimple{}&\Deightyonefiftyghzconstraintfullsimple{}& \alphagalconstraintfullsimple{} & \betagalconstraintfullsimple{} 
        & \Dfivehundredninetyfiveghzsyncpoissonconstraintfullsimple{} 
        & 3.0\%\\
        \bottomrule
    \end{tabular}
    \caption{The 95\% upper limits and estimated $1\sigma$ uncertainty on the tensor-to-scalar ratio $r$ from SPT-3G data for the BK18 and SPT-3G 1500\,deg$^2$ apodization masks. Parameters for the Galactic dust model and extragalactic radio galaxies are also reported. In the rightmost column, we report the PTE of the best-fit point from an F-distribution due to the uncertainty in the covariance matrix estimate. 
 }
                \label{tab:rmask}
\end{table*}
The auto- and cross-power spectra for the 95\,GHz, 150\,GHz, and 220\,GHz bands of SPT-3G using the BK18 mask are tabulated in Table~\ref{tab:bandpower_error}. 
The listed uncertainties include sample and noise variance and are based on the square root of the diagonal elements of the Gaussian covariance matrix.
In Figure~\ref{fig:bandpowers}, the bandpowers are plotted along with the best-fit model for Galactic and extragalactic foreground emission plus lensing $B$~modes.

\begin{figure}[t]
    \begin{center}
        \includegraphics[width=\linewidth]{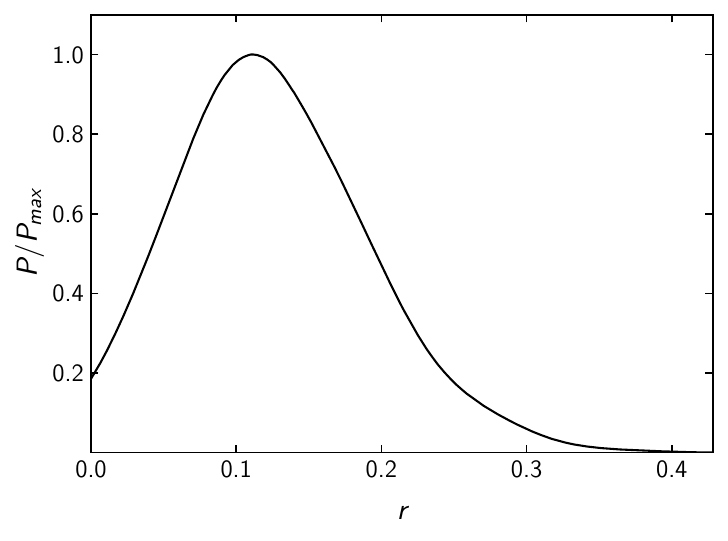}
        \caption{Posterior probability distribution on the tensor-to-scalar ratio, $r$, inferred from the measured SPT-3G bandpowers from the BK18 mask.}
        \label{fig:r1d}
    \end{center}
\end{figure}

\begin{figure}[t]
    \begin{center}
        \includegraphics[width=\linewidth]{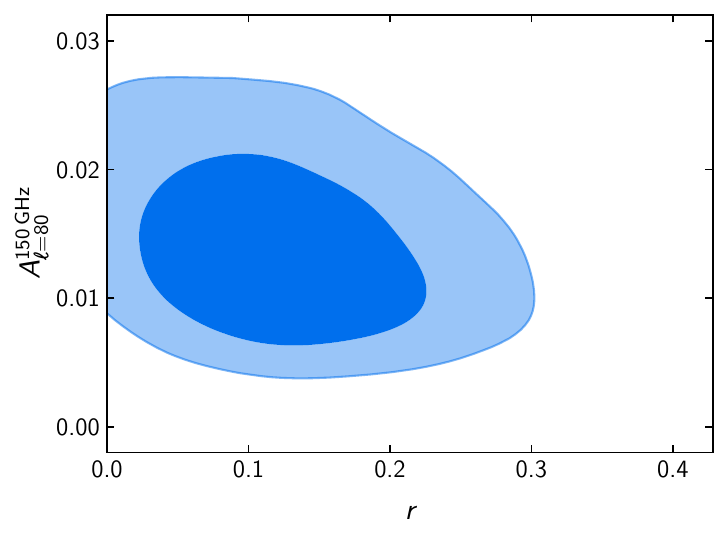}
        \caption{2D posterior probability distribution on the tensor-to-scalar ratio, $r$, and the Galactic dust amplitude at $\ell=80$ and 150\,GHz, $A_{\ell=80}^{150\,\mathrm{GHz}}$. The dark and light blue contours mark the 68\% and 95\% credible intervals, respectively. }
        \label{fig:rdust}
    \end{center}
\end{figure}

We fit these data to the model described in Section~\ref{section:model}, a 8-component model that includes $r$, three calibration terms, lensing $B$~modes, a Poisson term for radio galaxies, and  Galactic dust emission. We find the model provides a reasonable fit to the data with a PTE of 42\% at the best-fit point. The recovered posterior probability distribution of $r$ is shown in Figure~\ref{fig:r1d}, and the 2D posterior probability distribution on $r$ and the amplitude of Galactic dust is shown in Figure~\ref{fig:rdust}. As expected, we see a degeneracy between the amplitude of the dust power and $r$, as illustrated in Figure~\ref{fig:rdust}. The inferred power due to inflationary gravitational waves decreases as the modeled Galactic dust power increases. 

\begin{figure}[t]
    \begin{center}
        \includegraphics[width=\linewidth]{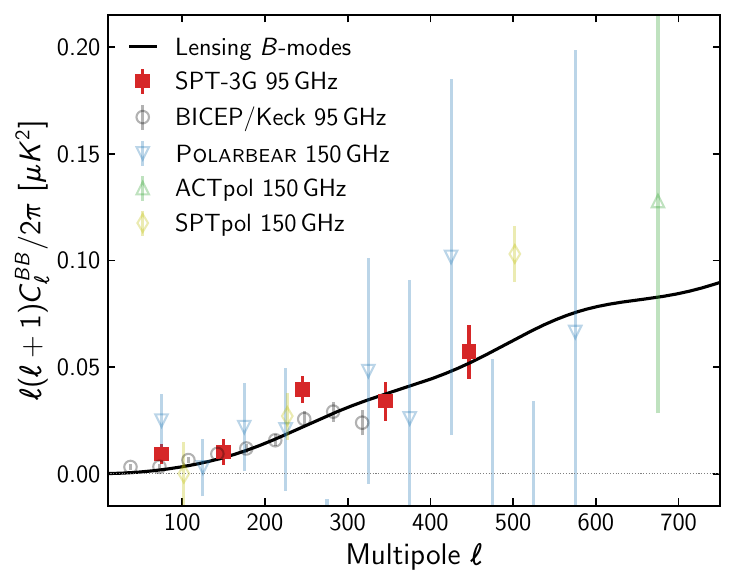}
        \caption{Ground-based $B$-mode measurement landscape. Data are from SPTpol \citep{sayre20}, ACT \citep{actpolbb}, \textsc{Polarbear} \citep{adachi22}, and BICEP/Keck \citep{bicepkeck21c}.}
        \label{fig:Bincontext}
    \end{center}
    \vspace{-15pt}
\end{figure}

We set a  95\% upper limit of  $r< \rlimitbksimple{}$, which is the best upper limit on $r$ from ground-based CMB $B$~modes outside of those from the BICEP/Keck family of experiments. Figure~\ref{fig:Bincontext} shows our  $B$-mode spectrum results alongside other recent ground-based measurements. 

The observed dust model parameters for the SPT-3G data using the BK18 mask are listed in Table~\ref{tab:rmask} and agree at approximately the $1\sigma$ level with the reported dust power in BK18. While this does not account for shared sample variance between the experiments, the SPT-3G measurement is instrumental noise limited.  The agreement serves as a cross-check of the BK18 result, as the SPT and BICEP/Keck experiments differ significantly in their instrument design and systematic effects. Key differences include the beam sizes ($\sim$1-2 arcmin for SPT-3G versus $\sim$30 arcmin for BICEP/Keck at 150\,GHz), optical designs, field of view, scan speed, approaches to ground shielding, and readout architecture. Despite these differences, the derived constraints on foreground parameters are consistent, demonstrating the robustness of the modeling and mitigation techniques across independent experimental platforms.  

\subsection{SPT-3G \texorpdfstring{1500\,deg$^2$}{1500 deg2} Survey Area}\label{section:fullfit}
We repeat the likelihood analysis using bandpowers derived from the full SPT-3G 1500\,deg$^2$ survey area. This yields a 95\% upper limit of $r < \rlimitfullsimple{}$, nearly identical to the constraint from the BK18 region. While the greater number of modes from the larger sky area reduces statistical uncertainty, interpreting this result requires accounting for the increased sample variance due to Galactic foregrounds and the added model complexity they introduce in this region.

The full SPT-3G footprint exhibits elevated levels of foreground power compared to the BK18 region. This discrepancy can be seen in the increase in amplitude of $A_{\ell=80}^{150\,\mathrm{GHz}}$ and $A^{\mathrm{95\,GHz,\,rg}}_{\ell=500}$ in Table~\ref{tab:rmask}, by-eye in Figure~\ref{fig:Emaps}, and through cross-correlation with \textsl{Planck} 353\,GHz polarization maps, a tracer of Galactic dust \citep{akrami2020planckfg}.  

\begin{figure}[t]
    \begin{center}
        \includegraphics[width=\linewidth]{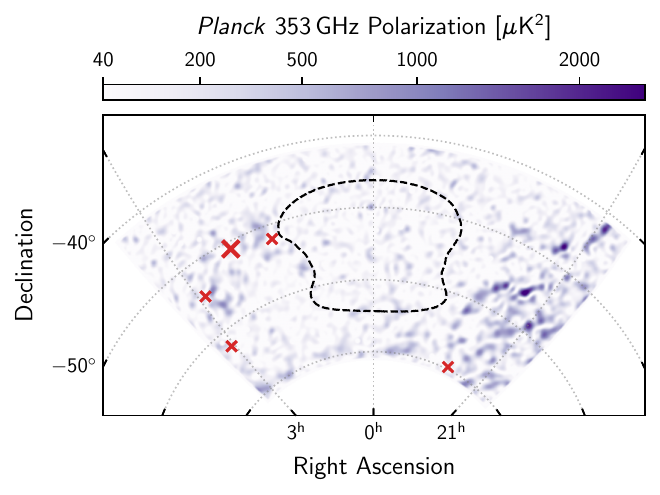}
        \caption{\textsl{Planck} 353\,GHz polarization ($P^2 = Q^2 + U^2$) in the SPT-3G survey area, bandpass-filtered over $\ell \in 30-200$ to highlight the large angular scales where the inflationary signal peaks. The black dashed line marks the half-power contribution threshold of the BK18 apodization mask. Red $\times$ symbols mark the locations of the five brightest polarized radio galaxies in the SPT-3G field, with the brightest one denoted with a larger $\times$. Both Galactic dust emission and polarized radio sources are more prominent outside the high-weight BK18 region, highlighting the increased foreground complexity across the full SPT-3G footprint.}
        \label{fig:foregrounds}
    \end{center}
    \vspace{-20pt}
\end{figure}

This contrast in foreground emission is further illustrated in Figure~\ref{fig:foregrounds}, which shows the \textsl{Planck} 353\,GHz polarization map across the SPT-3G survey area. The black dashed line marks the half-power contour of the BK18 apodization mask, highlighting how both Galactic dust and bright radio sources are more prominent outside this cleaner subregion. The \textsl{Planck} 353\,GHz polarization map shows significantly more diffuse dust emission outside the BK18 high-weight region, indicating higher and spatially variable Galactic dust power across the full SPT-3G field. Similarly, the five brightest radio galaxies in the SPT-3G survey area, marked by red $\times$ symbols, are located outside the high-weight region of the BK18 mask. The brightest of these (marked by the larger $\times$) is located where the mask assigns only 40\% weight and has four times more flux than the next brightest source, making it the dominant contributor from radio galaxies in our fit. 

This elevated and more complex foreground power is not well described by the simple model presented in Section~\ref{section:model}. While the model performs adequately over the BK18 region (PTE = 42\%), it yields a significantly degraded fit over the full footprint (PTE = 3.0\%). Given that the model succeeds in the cleaner region and the excess power across the full field has a dust-like SED, we consider more complex Galactic foregrounds the most likely explanation. For instance, spatial variations in the dust SED, such as those reported by \textsc{Spider} \citep{ade25}, are not captured by the assumptions of the current model.

Although the fit is poor, the resulting upper limit on $r$ is likely conservative. Excess unmodeled foreground power at $\ell$ = 80 biases $r$ upward rather than artificially suppressing it, as excess power not well fit by the foreground model will be interpreted as the inflationary signal. We explored this effect by implementing more flexible foreground models in the likelihood, e.g., ones allowing for different power or SEDs as a function of multipole, and found that the inferred upper limits on $r$ decreased. Adding more degrees of freedom to the foreground modeling generically reduces the bias and tightens the constraint. While we cannot fully rule out residual systematics or covariance misestimation, either would similarly add power and bias $r$ high. The $1\sigma$ uncertainty on $r$ is also larger than in the BK18 region, as the additional cosmic variance penalty from the higher dust power increases more rapidly than the number of modes gained by observing more sky.

This tradeoff highlights the importance of field selection in future survey design. Expanding directly around the BICEP/Keck field will likely require more sophisticated foreground modeling or subtraction techniques to fully realize the sensitivity gains of next-generation instruments.

\section{Conclusion}\label{section:conclusions}
 
We present new measurements of the $B$-mode power spectrum using two years of SPT-3G data. The bandpowers are reported over the angular multipole range 32 $< \ell <$ 502 for 6 frequency combinations: $95\times95$\,GHz, $95\times150$\,GHz, $95\times220$\,GHz, $150\times150$\,GHz, $150\times220$\,GHz, and $220\times220$\,GHz.  
We compare these data to bandpowers from other experiments in Figure~\ref{fig:Bincontext}. 
We fit the observed $B$-mode power to a model consisting of a combination of lensing-induced $B$~modes, extragalactic foregrounds, and Galactic dust. We place a 95$\%$ upper limit on the tensor-to-scalar ratio of $r$ $<$ \rlimitbksimple{}. 

This work describes several key advancements in the search for inflationary $B$~modes. First, the data for this work overlaps with the BICEP3 field, the proposed CMB-S4 patch, and the Simons Observatory survey region. We observe consistent levels of Galactic dust emission as BK18 over the same sky area, but with an independent analysis pipeline and different instrumental systematics. The higher levels of Galactic dust emission we measure at the edges of the BICEP/Keck field may inform future survey design. 

Second, these maps are well suited for future work in cross-correlation with other experiments with overlapping sky area such as the BICEP/Keck surveys, Simons Observatory, or CMB-S4.  These analyses can incorporate higher frequency bands to improve constraints on Galactic dust, as well as for validating component separation algorithms of both astrophysical signals and instrumental systematics across a wide patch of sky. Techniques to improve the reconstruction of large-angular scale modes, like in this work, are also relevant to gravitational lensing measurements. For the $EB$ lensing estimator in particular \citep{namikawa2014bias}, the lensing signal peaks at $\ell \sim$500 and the polarized noise performance at these scales is very important. Gravitational lensing measurements are critical to future constraints on parameters such as the sum of neutrino masses that affect the growth of large scale structure, and also to delens future primordial gravitational wave searches.

Third, while investigating SPT-3G low-frequency noise for this analysis, we determined that polarized atmospheric emission was the limiting noise source for SPT-3G on the angular scales relevant to inflationary constraints. We demonstrate an unbiased method to remove this noise from the data utilizing co-pointing detectors of different frequencies and the fixed spectral scaling of polarized atmospheric emission. Polarized atmospheric noise has been observed in experiments at both the South Pole and in Chile \citep[e.g.,][]{takakura19, li23b, coerver24}. However, this is the first CMB analysis that explicitly mitigates polarized atmospheric noise. This analysis serves as a demonstration that it is possible to produce precise measurements of large-scale polarization anisotropy with a large aperture ground-based telescope.

With two years of data, we have presented large-angular-scale noise performance with SPT-3G and the second-best ground-based $B$-mode constraint on $r$. Another three years of SPT-3G data on the 1500\,deg$^2$ patch (from 2021--2023) already exist, and at least two more years (2025--2026) are planned. If the noise continues to integrate down without being limited by instrumental systematic effects, the noise variance on $r$ will be reduced by a factor of 2.5 with existing data taken through 2023 and by a factor of 3.5 with data to be taken through 2026.
 
\acknowledgments
The South Pole Telescope program is supported by the National Science Foundation (NSF) through awards OPP-1852617 and OPP-2332483. Partial support is also provided by the Kavli Institute of Cosmological Physics at the University of Chicago. Argonne National Laboratory’s work was supported by the U.S. Department of Energy, Office of High Energy Physics, under contract DE-AC02-06CH11357. The UC Davis group acknowledges support from Michael and Ester Vaida. Work at the Fermi National Accelerator Laboratory (Fermilab), a U.S. Department of Energy, Office of Science, Office of High Energy Physics HEP User Facility, is managed by Fermi Forward Discovery Group, LLC, acting under Contract No. 89243024CSC000002. The Melbourne authors acknowledge support from the Australian Research Council’s Discovery Project scheme (No. DP210102386). The Paris group has received funding from the European Research Council (ERC) under the European Union’s Horizon 2020 research and innovation program (grant agreement No 101001897), and funding from the Centre National d’Etudes Spatiales. The SLAC group is supported in part by the Department of Energy at SLAC National Accelerator Laboratory, under contract DE-AC02-76SF00515. This research was done using services provided by
the OSG Consortium, which is supported by
the National Science Foundation awards 030508 and 1836650. Support for this work for J.Z. was provided by NASA through the NASA Hubble Fellowship grant HF2-51500 awarded by the Space Telescope Science Institute, which is operated by the Association of Universities for Research in Astronomy, Inc., for NASA, under contract NAS5-26555.
  
\bibliography{biblio}
\bibliographystyle{aasjournal}

\end{document}